\documentclass[showpacs,prl,aps]{revtex4}

\usepackage{graphicx}
\usepackage{dcolumn}
\usepackage{bm}
\begin{document}
\title{Comment on ``Model for Heat Conduction in Nanofluids''}
\author{Sorin Bastea}
\email{bastea2@llnl.gov}
\affiliation{Lawrence Livermore National Laboratory, P.O. BOX 808, Livermore, CA 94550}
\pacs{44.10.+i, 44.35.+c}
\maketitle
In a recent letter \cite{kpk04}, Kumar et al. outline a model for thermal transport 
in nanofluids that purports to explain the anomalous thermal 
conductivity enhancements observed experimentally. Since this effect is expected to 
have significant technological consequences and may already be relevant for certain 
applications \cite{gpj88} this is an important exercise. Unfortunately, the model of 
Kumar et al. falls short on several conceptual counts.

The authors start by 
considering the case of fixed solid particles and postulate the existence of 
``two parallel paths of heat flow'', one through the suspending liquid and the other 
through the dispersed solid. This is a questionable assumption, 
particularly for the extremely dilute suspensions against which they wish to test their model. 
By pursuing this argument along an approximate derivation Kumar et al. produce a 
relation for the effective thermal conductivity of the nanofluid, Eq. 7 of \cite{kpk04}, 
which includes a dependence on the radiuses of the solid particles, $r_p$, and 
liquid molecules, $r_m$:
\begin{eqnarray}
k_{eff}=k_m\left[1+\frac{k_p\epsilon r_m}{k_m(1-\epsilon)r_p}\right]
\end{eqnarray}
,where $k_p$ and $k_m$ are the thermal conductivities of the solid and liquid 
respectively, and $\epsilon$ is the volume fraction of the dispersed solid particles. One problem 
with the above relation is immediately apparent if we notice that at fixed 
$\epsilon$, $r_m/r_p \rightarrow 0$ implies $k_{eff}/k_m\rightarrow 1$, i.e. 
the solid inclusions have no effect on the thermal conductivity if they are much larger than the 
liquid molecules. In fact, this situation is rigorously described by theories that treat the 
liquid as a continuum and where accurate, widely accepted results are available for all 
$\epsilon$ \cite{ct90}. The dependence of Eq. 1 on $k_p/k_m$ is very different 
from these results even at $r_m/r_p>0$. For example, if the solid clusters 
are much less thermally conducting than the liquid, 
$k_p/k_m\rightarrow 0$, Eq. 1 predicts $k_{eff}/k_m\rightarrow 1$, 
while if they are much more conducting, $k_p/k_m\rightarrow \infty$, it yields $k_{eff}/k_m\rightarrow \infty$, 
both independently of $\epsilon$ (and $r_m/r_p$), and both of which can only be deemed incorrect. 

A dependence of $k_{eff}$ on $r_p$ can arise 
for example due to liquid-solid interface thermal resistance \cite{tr95}, even when the liquid 
can be treated as a continuum. Furthermore, as $r_p/r_m$ decreases the effective volume fraction 
of the solid particles, 
e.g. approximated by $\epsilon_{eff}=\epsilon(1+r_m/r_p)^3$ for $\epsilon\ll 1$,
should also perhaps replace $\epsilon$ in conductivity estimates, introducing an additional 
dependence on $r_p$. However, Eq. 1 captures no such effects and fails some simple 
tests. 

Kumar et al. consider next the effect of the solid particles motion. 
To this end they invoke kinetic theory and introduce $c\bar u_p$ as the 
``thermal conductivity of the particle'', where $\bar u_p$ is an ``average particle 
velocity''. Their complete thermal conductivity model is obtained by replacing $k_p$ in Eq. 1 
with $c\bar u_p$. This is quite problematic since it has the effect of eliminating $k_p$ as a 
parameter in determining the nanofluid thermal conductivity, and thus makes the previous analysis 
for fixed 
solid clusters rather futile. For example, if $\bar u_p \rightarrow 0$, .e.g. the suspending 
liquid is frozen, $k_{eff}\rightarrow k_m$, i.e. the solid inclusions have no effect on 
$k_{eff}$ irrespective 
of $k_p$ and $\epsilon$, when this case should reduce to the fixed particles 
problem. 

In fact, $c\bar u_p$ cannot be interpreted at all as the thermal conductivity of a solid 
particle. By the authors own kinetic theory arguments 
$k\prime=c\bar u_p$ is the thermal conductivity of a dilute gas of particles that possess 
internal energy \cite{cc}. In principle, a quantity like $k\prime$ 
could provide an estimate for direct, Brownian motion transport of heat by the 
solid clusters, but it should not 
supplant $k_p$, the thermal conductivity of the solid. Unfortunately, 
the $k\prime$ calculated in \cite{kpk04} is inadequate even for this purpose. It is not 
clear why the authors definition of $\bar u_p$ would be appropriate in the given context, 
while their 
estimate of the ``constant'' $c\propto n l c_v$ (which 
is not dimensionless as conveyed in the paper), with $n$ - number density 
of the solid particles, $l$ - their ``mean-free path'' and $c_v$ - heat capacity of a 
solid particle, apparently assumes $l\propto 1/nd^2_p$, $d_p=2 r_p$, which only takes into account 
the rare collisions between the solid particles and ignores the effect of the liquid. 
A more reasonable estimate for $k\prime$ is $k\prime \simeq nDc_v$, with 
$D=k_B T/3\pi\eta d_p$ - Stokes-Einstein relation. This yields for $d_p=10nm$ gold particles 
in water at normal conditions and $\epsilon=1\%$, $k\prime/k_m\simeq 10^{-6}$. 
Therefore, the direct Brownian motion contribution to thermal transport can be safely 
ignored for nanofluids. Finally, the theoretical values plotted by the authors in 
their Fig. 4 are off by orders of magnitude from their own formula Eq. 10, purportedly 
used to calculate them.  

This work was performed under the auspices of the U. S. Department of Energy by 
University of California Lawrence Livermore National Laboratory under Contract 
No. W-7405-Eng-48.

\end{document}